\documentstyle{elsart}
\def\MN{\em Mon.~Not.~R.A.S.} 
\def\APJ{\em Astrophys.~J.} 
\def\APJS{\em Astrophys.~J.~S.~S.}

\def\PLB{{\em Phys. Lett.}  B}
\def\Na{{\em Nature}}
\def\PLA{{\em Phys. Lett.}  A}
\def\PRL{\em Phys. Rev. Lett.}
\def\PRD{{\em Phys. Rev.} D}

\def\AA{ $A.$\&$A.$}
\def\SCI{\em Science}

\def\be{\begin{equation}}
\def\ee{\end{equation}}
\def\bea{\begin{eqnarray}}
\def\eea{\end{eqnarray}}
\def\mincir{\raise -2.truept\hbox{\rlap{\hbox{$\sim$}}\raise5.truept
\hbox{$<$}\ }}
\def\magcir{\raise -4.truept\hbox{\rlap{\hbox{$\sim$}}\raise5.truept
\hbox{$>$}\ }}

\begin{document}
\begin{frontmatter}
\title{MIXED MODELS WITH $n>1$ AND LARGE SCALE STRUCTURE
CONSTRAINTS}

\author{ S.A. Bonometto }

\address{Dipartimento di Fisica dell'Universit\`a \& INFN sez. di Milano\\
 Via Celoria 16,\\ I20133 Milano, ITALY\\
e--mail: bonometto@mi.infn.it\\
}

\author{ E. Pierpaoli }

\address{SISSA -- International School for Advanced Studies,
Via Beirut 2/4, \\I34013 Trieste, ITALY\\
e--mail: pierpa@sissa.it\\
}

\begin{abstract}
Recent data on CBR anisotropies show a Doppler peak higher than expected in 
CDM cosmological models, if the spectral index $n=1$. However, CDM and LCDM
models with $n>1$ can hardly be consistent with LSS data. Mixed models,
instead, whose transfer function is naturally steeper because of 
free--streaming in the hot component, may become consistent with data if 
$n>1$,  when $\Omega_h$ is large. This is confirmed
by our detailed analysis, extended both to models with a hot component whose 
momentum space distribution had a thermal origin (like massive neutrinos), 
and to models with a non--cold component arising from heavier particle decay.
In this work we systematically search models which fulfill all constraints 
which can be implemented at the linear level. We find that a stringent 
{\sl linear} constraint arises from fitting the {\sl extra--power} parameter 
$\Gamma$. Other significant constraints arise comparing the expected 
abundances of galaxy clusters and high--$z$ systems with observational data. 
If low values of $\Gamma$ are permitted, mixed models with $1 \leq n \leq
1.4$
can have up to $\sim 45\, \%$ of non--cold component, without violating any
further {\sl linear} constraint. Keeping to models with $\Gamma \geq 0.13$, a
suitable part of the space parameter still allows up to $\sim 30\, \%$ of
hot component (it is worth outlining that our
stringent criteria allow only models with $0.10 \mincir \Omega_h \mincir 0.16$,
if $n \leq 1$). Spectra with $n>1$ are briefly discussed within the frame of
inflationary theories. We also outline that models with such large non--cold
component would ease the solution of the so--called {\sl baryon catastrophe} in
galaxy clusters. 
\vskip 0.2 truecm

PACS: 95.35; 98.80; 98.65.Dx; 98.62.L

\end{abstract}

\begin{keyword} cosmology:theory,
 dark matter, 
 large scale structure of the Universe.  
\end{keyword}
\end{frontmatter}

\section{Introduction}

We need three kinds of ingredients to define a cosmological model:
the background metric, the substance mix and the form of early departures
from homogeneity. Background metric features are set 
by the density parameter $\Omega =\rho/\rho_{cr}$ and the
Hubble parameter $H = h\, 100\, $km$\, $s$^{-1}$Mpc$^{-1}$ 
($\rho$ is the average density and $\rho_{cr} = 3H^2/8\pi G $);
 the {\sl cosmic substance} is fixed 
by partial density parameters, like $\Omega_b = \rho_b/\rho_{cr}$
for baryons,  $\Omega_c = \rho_c/\rho_{cr}$ for cold--dark--matter
(CDM), or $\Lambda$ for vacuum; finally,
early  deviations from homogeneity  are described by the
amplitude and the spectral index $n$ of the initial fluctuation
spectrum.

However, while early spectral features essentially do not depend
on background and substance parameters, the  linear evolution
of the spectrum does and this is usually expressed through transfer functions
${\cal T}_z$$(k)$ ($k = 2\pi/L$, $L$ is the comoving length--scale),
which give the ratio between the spectrum at a redshift $z$ and its
initial value.

In this paper we shall restrict ourselves to spatially flat models, with no
residual vacuum energy, and focus on the interplay between
the nature of the cosmic substance (yielding ${\cal T}_z$)
and primeval spectral features, performing a systematic analysis of the amount
and nature of hot dark matter (HDM), allowed by observational
constraints, once we take the primeval spectral index
(see below) in the range $ 1 \leq n \leq 1.4$. 

In this work we shall however consider only large scale structure 
(LSS) features obtainable from linear theory, together with data 
on cosmic background radiation (CBR) anisotropies, which also 
refer to an epoch when fluctuations were still linear.
The CBR temperature is almost direction independent
and, once the dipole term (which accounts for the motion
of Earth in respect to CBR) is subtracted, its fluctuations  
$\Delta T(\theta,\phi)/T = \sum_{l,m} a_{lm} Y_{lm}(\theta,\phi)
$$\mincir 10^{-5}$. Since we expect that deviations from a Gaussian behaviour 
are substantially absent at these stages, anisotropy features are 
suitably expressed through the quantities $C_l = \langle |a_{lm}|^2 
\rangle $, for $l \geq 2$. 
As we shall see, CBR and LSS linear constraints already cause a severe
selection among models and models to be used in
numerical simulations should be submitted first to {\sl linear } tests.

Let us now consider recent data on CBR anisotropies.
As usual, we shall assume that the primeval spectrum reads
$$
P_{\Psi} (k) = A_{\Psi} x^3 (x k)^{n-4}
\eqno (1.1)
$$
($x = x_o - x_{rec}$  is the distance from the recombination band,
as $x_o$ is the present horizon radius). It is then known (see,
$e.g.$, Ma $\&$ Bertschinger 1996)  that
$$ 
C_l = (4~ \pi)^2 \int dk\, k^2  P_{\Psi}(k) ~ \Delta^2_l (k, x_o)  
\eqno (1.2)
$$
with
$$
\Delta_l(k, x_o) = \int_0^{x_o} dx  S (k,x) j_l [(x_o - x)k]  
\eqno (1.3)
$$  
(Seljak \& Zaldarriaga 1996).
In this general expression, the source function $S (k,x)$ contains 
contributions originating both at the last scattering band and 
along the path from it to us. For $\Omega=1$ and low $l$ (large scales $L$),
the most relevant effect is Sachs--Wolfe, and the simple form
$$
C_l \simeq (2\pi/3)^2 A_{\Psi} \int (dk/k) (x_o k)^{n-1} j_l^2 (x_o k) 
\eqno (1.4)
$$
can be used, which does not depend on the substance characteristics, 
but only on $n$. Although this approximation holds for rather small
$l$ only, in most cases the nature of the substance does not drastically affect
the $l$ dependence of $C_l$ up to the so--called Doppler peak
($l \sim 200$) (Dodelson et al. 1996).

Using eq.~(1.4) and measuring $C_l$ up to $l \sim 30$, the COBE collaboration 
(Bennet et al. 1996) could estimate the spectral index and found $n=1.2 \pm 
0.3$. Therefore, COBE data already favour $n >1$, although they are fully 
compatible with $n=1$.

More recent measurement (Netterfield et al., 1997), extending up to $l \simeq 
400$ (corresponding to angles $\sim 0.5^o$), explored the 
doppler peak. The detected values seem to exclude models with 
low $\Omega$ values (Hancock et al. 1998), and agree with CDM and $\Omega=1$ 
only for $h = 0.3$ (Lineweaver 1997, Lineweaver $\&$ Barbosa 1998). 
An alternative possibility to fit such data is that $n > 1$. 

Although these results obviously need further confirm, they 
seem to tell us that inflation generated a spectrum $P(k)$ with $n > 1$.

There are two main reasons why this is seen as a problem by some
cosmologists. In the first place, the generic prediction of inflation is that 
$n \simeq 1$. Section 4 will be devoted to a brief discussion on a class
of inflationary theories which however predict $n >1$.
Furthermore, CDM fit with LSS improves for $n < 1$.

Another way to improve the fit of models with LSS consists into assuming
that DM has also a hot component. The basic philosophy of mixed dark matter 
(MDM) models, since the first computation of their transfer function 
(Bonometto $\&$ Valdarnini, 1984; see also Achilli et al. 1985; 
 Valdarnini $\&$  Bonometto 1985), is that the unexplored non--interacting 
world (gravitation apart) can be quite complicated and
probably contains several particle components. If one of them becomes
non--relativistic when the scale $\bar L$ enters the horizon, then $\cal T$ 
is damped over all scales $L \mincir \bar L$. The detailed expression 
of $\cal T$ depends on the DM mix and the effect cannot be
neglected if such component gives a substantial contribution to the
overall density and $\bar L$ exceeds galaxy scales.

An attempt was also performed to consider $n>1$ in association to
mixed DM (Lucchin et al. 1996), trying to balance
opposite departures from standard CDM. However, the mix studied
by these authors produces too many clusters, as was outlined by 
Borgani et al. (1996). Liddle et al. (1996) considered models with
$n > 1$, in association with high $\Omega_h$ and possibly
low $h$ values, but they saw the same problems outlined by Borgani et al (1996)
and did not indicate an area of fit with data. If it is confirmed 
that $n>1$, however, there seems to be no alternative to try to follow
such pattern. 

Owing to that, we tested a wide range of DM mixtures in association
with values of $n$ between 1 and 1.4. We found that the above difficulties 
can be overcame for a wide set of such models. Our analysis was 
restricted to the {\sl linear} constraints described in sec.~3 and to 
models with $\Omega = 1$, $\Omega_b = 0.1$, $\Lambda = 0$ and $h = 0.5$.
Models were parametrized by $\Omega_h$ (the density parameter of the 
non--cold component) and $d = z_{eq}/z_{der}$; here we defined $z_{eq} \equiv
10^4$; this would be the redshift when
the photon density reaches the baryon+DM density, if
all DM is CDM and $\theta = T/2.733\, {\rm K} = 1$. The derelativization 
redshift 
$z_{der}$, instead, occurs when the average particle momentum $\langle p 
\rangle = m$ (particle mass). 

A particle model allowing a continuous
selection of $[\Omega_h,d]$ pairs is obtained if heavier particles ($X$)
decay
into lighter massive particles ($v$), whose phase--space distribution
turns out to be quite different from {\sl usual} massive $\nu$'s. Particle 
models leading to such hot DM component were discussed in several papers 
($e.g.$, Bonometto, Gabbiani $\&$ Masiero 1994, Ghizzardi 
$\&$ Bonometto 1996) and the evolution 
of density fluctuations in such context was discussed by Pierpaoli $\&$ 
Bonometto (1995) and Pierpaoli et al. (1996). Hereafter we shall distinguish
these models from models with phase--space particle distribution
of thermal origin, calling {\sl volatile} their non--cold component.
However we shall show that also a large deal of models based on massive $\nu$'s
are consistent with linear constraints.

Owing to the peculiar phase--space distribution of volatile DM and to the 
actual energy dependence of fluctuation amplitudes in such models,
public algorithms cannot be used to evaluate $\cal T$.
Here we shall make use of a flexible algorithm that we developed also
to this aim. A discussion of phase--space distributions,
fluctuation evolution and constraints arising from particle
kinematics is given in sec.~2 and
in Appendix A. More details on the algorithm are given in Appendix B.
As already mentioned, sections 3 and 4 deal with
{\sl linear} LSS constraints and inflationary theories yielding $n>1$,
respectively. Section 5 will be devoted to the analysis of results
concerning acceptable models. We provide some plots showing the allowed
area in the parameter plane and some sample transferred spectra.
Furthermore, for parameter choices consistent with hot DM 
($e.g.$ massive neutrinos), we compare the transferred 
spectra, finding small (expected) discrepancies between volatile models 
and those with massive $\nu$'s. Section 6 is devoted to a final discussion
and to conclusions.

\section{Parameter space limitation}

The {\sl substance} of mixed models is made of various components,
which can be classified according to their behaviour at the time when
galactic mass--scales enter the horizon. Particles which become 
non--relativistic before then, are said {\sl cold}.
There can be one or more of such components,
indistinguishable through astrophysical measures.
{\sl Tepid} or {\sl warm}  components were sometimes considered
(see, $e.g.$ Colombi et al. 1996; Pierpaoli et al., 1998), made of
particles which become non--relativistic
when small galactic scales ($\sim 10^{8} M_\odot$) have entered the
horizon, but large galactic scales ($\sim 10^{12} M_\odot$) still have not.
Finally, {\sl hot} component(s) are made of particles which become 
non--relativistic after the latter scale has entered the horizon.

Neutrinos, if massive, are a typical hot component. They were
dynamically coupled to radiation down to a temperature $T_{\nu,dg} 
\sim 900\, $keV. If their mass $m \ll T_{\nu,dg}$, their number
density, at any later time, is $n_\nu = (3\zeta(3)/4\pi^2) g_\nu T_\nu^3$
(after electron annihilation $T_\nu = T_{\nu,dg} [a_{\nu,dg}/a(t)] 
= (4/11)^{1/3} T$, where $T$  is radiation temperature) and 
their momentum distribution (normalized to unity) reads
$$
\Phi_\nu (p,t) = {2 \over 3 \zeta(3)}
{(p^2/T_\nu^3) \over \exp[p/T_\nu(t)] + 1}
\eqno (2.1)
$$
also when $p \ll m$. Henceforth, when $T \ll m$, their distribution
is not {\sl thermal}, although its shape was originated in
thermal equilibrium. Notice that, for high $p$, $\Phi_\nu$ is
cut off as $\exp(-p/T_\nu)$.

Using the distribution (2.1) we can evaluate
$$
\langle p \rangle = (7 \pi^4 / 180 \zeta(3) ) T_\nu = 3.152\, T_\nu~.
\eqno (2.2)
$$
Accordingly, $\langle p \rangle = m$ when $T_\nu = 0.317\, m$ and, for 
massive $\nu$'s, 
$$
d = 5.2972 \cdot 4h^2/(m/{\rm eV})~,
\eqno (2.3)
$$
while
$$
\Omega_h = 2.1437 \cdot 10^{-2} g_\nu 
(m/{\rm eV}) \theta^3/4h^2 
\eqno (2.4)
$$

Hot components can also originate in the decay of a heavy particle $X$.
Let $N_{X,dg}$ be the comoving number density of $X$'s at their
decoupling, taking place at some early time $t_{dg}$, well after they
became non--relativistic, and let $m_X$ be their mass. At $t \gg t_{dg}$
their comoving number density reads:
$$
N_X (t) = N_{X,dg} \exp[-(t-t_{dg})/\tau_{dy}]
\eqno (2.5)
$$
with $t_{dg} \ll \tau_{dy}$ (decay time). We shall assume a two--body
decay process, as is more likely for dynamical reasons:
$$
X \to v + \phi
\eqno (2.6)
$$
The decay gives rise to a light ({\sl volatile}) particle $v$, of mass $m 
\ll m_X$, and to a massless particle $\phi$. It can be either a photon 
($\gamma$) or a sterile scalar, as is expected to exist in theories where a 
global invariance is broken below a suitable energy scale (examples of
such particles are {\sl familons} and {\sl majorons}). 
We shall show that the latter case (sterile
scalar) is potentially more significant.

Once the decay process is over, the volatile momentum distribution reads:
$$
\Phi_v (p,t) = 2 (Q/p) \exp(-Q) ~~~{\rm where} ~~~ Q = p^2/ {\tilde p}^2
\eqno (2.7)
$$
and
$$
{\tilde p} = (m_X/2) [a_{dy}/a(t)]
\eqno (2.7a)
$$
provided that $X$'s, before they decay, never attain a density exceeding 
relativistic components. As is shown in Appendix A, this
is avoided if the condition
$$
\Omega_h d \ll [1+2/(1+g_\nu/8.8)]^{-1} ~~~~~~{\rm for~}\gamma{~decay}.
\eqno (2.8)
$$
is fulfilled. Here $g_\nu$ refers to massless $\nu$'s; $e.g.$, for
$g_\nu = 6$, eq.~(2.14) yields $\Omega_h d < 0.46$. If matter dominance 
occurs, the distribution (2.7) is distorted at small $p$, $i.e.$ for particles
originating in early decays. In this case the distribution can still be 
evaluated numerically, but no simple analytical expression can be given.
Particles with large $p$, instead, were born in decays occurring
when radiation dominance was recovered and, therefore, the distribution 
is however cut off as $\exp (- p^2/{\tilde p}^2)$. 
The cut--off is therefore sharper than in the neutrino case.

Using the distribution (2.7), it is easy to see that the average
$$
\langle p \rangle = (\sqrt{\pi} / 4) m_X a_{dy}/a(t) =  
{\tilde p} \sqrt{\pi} / 2 ~,
\eqno (2.9)
$$
according to eq.~(2.7a). $v$'s will therefore become non relativistic
when ${\tilde p} = 2\, m/\sqrt{\pi}$; henceforth
$$
{\tilde p} = (2md/\sqrt{\pi})(z/z_{eq})
\eqno (2.10)
$$
(let us recall that we defined $z_{eq} \equiv 10^4$). Eq.~(2.10)
can be conveniently used in the distribution (2.7), instead of eq.~(2.7a),
as involves model parameters ($m$ and $d$), instead of
early particle features ($m_X$ and $\tau_{dy}$), which should be
further elaborated in order to work out $a_{dy} = a(\tau_{dy})$.

Let us now discuss the decay patterns in more detail.
First of all, if $\phi \equiv \gamma$, the decay should occur at a 
redshift $z \magcir 10^7$, so to avoid distorsions of CBR 
spectrum (see, $e.g.$ Burigana et al., 1991). 
Two further possibilities should then be considered.

If the decay takes place before big--bang nucleosynthesis (BBNS), the
energy density of $v$'s is limited by the requirement that BBNS yields
are consistent with observed nuclide abundances; henceforth the number of 
relativistic spin states $g_\nu \mincir 7$ (see, $e.g.$ Olive et al.
1997).
Therefore, the actual limit on the energy density of $v$'s depends
on the number of relativistic $\nu$'s at BBNS. If some $\nu$ has a
mass exceeding $\sim 1\, $MeV (particle data surely allow it for
$\nu_{\tau}$) and decays before decoupling, $v$'s can {\sl
occupy} its place during BBNS. 
As far as cosmology is concerned, differences
between such volatile DM and ordinary hot DM would arise from the following 
3 kinematic reasons:
(i) the different phase--space distribution of $v$'s in respect
to massive $\nu$'s; (ii) the different energy dependence of fluctuation
amplitudes; (iii) the allowed continuous range of HDM derelativization 
redshift;
in the neutrino case, for a given HDM density, the derelativization
redshifts are ``quantized'' by the discreteness of $g_\nu$. 

By evaluating the transfer function for volatile models, we shall 
see that the points (i) and (ii) have a modest impact
(see Table I in sec.~5). Consequences of
(iii) would be important only if other cosmological parameters, like
$\Omega_b$ and $h$, were very well known. Altogether, therefore, the main
differences  induced by the above 3 points
concern particle aspects. For example, $v$'s number density
can be significantly smaller than $\nu$'s and their masses can be
much greater. In spite of that, their density and
derelativization redshift $z_{der}$ can be equal and LSS effects
are quite similar.

Let then still be $\phi \equiv \gamma$, but let us assume that the decay 
occurs after BBNS; also in this case, there
are constraints related to BBNS. In fact, decays produce an equal 
amount of $v$'s and $\gamma$'s. The present CBR temperature normalizes
$\gamma$ abundance {\sl after} the thermalization of decay photons.
Accordingly, a large $v$ density implies a small $\gamma$ and massless $\nu$
density {\sl before} the decay. BBNS would therefore take place almost at the
{\sl usual} temperature, but earlier in time. Neutron decay would be
therefore allowed a smaller time to work and neutron abundance,
at the opening of the so--called {\sl Deuterium Bottleneck}, would be greater.
From this qualitative framework quantitative limits to the final $v$ 
abundance allowed by BBNS can be derived. We shall not further
detail this point here and will only consider a softer
limit, ensuing from the obvious requirement that the final
$v$ density shall be smaller than $\gamma$ density, as the latter
one has a ``contribution'' from ``primeval'' $\gamma$'s. 
As is shown in Appendix A, for a model with given $\Omega_h$ and $d$, at
redshifts $z \gg z_{der}$, the following relation holds:
$$
\rho_h/\rho_\gamma = \Omega_h d~~~.
\eqno (2.11)
$$
Henceforth, in this case, it must be
$$
\Omega_h d \ll 1 ~~~~~~~~~~~ {\rm (\gamma~decay)}
\eqno (2.12)
$$
Therefore, according to eq.(2.8), this case never implies primeval
temporary matter--dominated expansion, at $t < {\tau}_{dy}$.
It should be also outlined that these models can have quite a low
massless neutrino density, as $X$ decay increases $\gamma $ temperature,
but does not act on $\nu$ background. In Appendix A we show that
$$
T_\nu / T_\gamma = (4/11)^{1/3} (1-\Omega_h d) ~~~~~~~~~~~ {\rm (\gamma~decay)}
\eqno (2.13)
$$
and  a relevant variation of the massless sterile
component (SMLC hereafter) can have a significant impact on ${\cal T}_z$.

On the contrary, if $\phi$'s are sterile scalars and the decay takes place
well after BBNS, constraints are not so stringent. $\phi$'s will
behave just as massless $\nu$'s and these models can be characterized
by a high SMLC density, which can also have
a major impact in shaping the present LSS.

Therefore, significant constraints on the models can arise from
this effect, which, however, cannot be discriminated {\sl a priori}.
Let us recall that, in the absence of $X$ decay, the ratio
$\rho_\nu / \rho_\gamma = 0.68132(g_\nu/6) \equiv w_o$. $X$ decay 
modifies it, turning $g_\nu$ into an effective value
$$
g_{\nu,eff} = g_\nu + (16/7)(11/4)^{4/3}\Omega_h d ~
\eqno (2.14)
$$
(see appendix A). In this case, no matter dominance occurs before $\tau_{dy}$
if
$$
\Omega_h d < (1+w_o)/2 ~~~~~~~{\rm (for~sterile~decay)}
\eqno (2.15)
$$
and $\phi$'s contribution to the relativistic component
lowers the {\sl equivalence} redshift. When
when $ d(1-\Omega_h) > 1+w_o$, also $v$'s are still relativistic
at equivalence. Accordingly, the equivalence occurs at either
$$
{\bar z_{eq}} = {4h^{2} \over \theta^4}
{10^4 \over 1+w_o+\Omega_h d}  ~~~~ {\rm or} ~~~~
{\bar z_{eq}} = {4h^{2} \over \theta^4} 
{10^4 (1-\Omega_h) \over 1+w_o+2\Omega_h d}
\eqno (2.15)
$$
in the former and latter case, respectively.

Mixed models involving a large volatile fraction, with late derelativization,
are therefore allowed only within this scenario. In what follows
we shall systematically analyze a large deal of mixed models, also
with large $\Omega_h$ and $d$, consistent with this last picture, but
we shall find that viable models with $\Omega_h d > 1$ are not so frequent.

\section{Linear constraints}

Model parameters can be constrained from particle physics and/or from LSS.
In this work we analyze a number of the latter constraints,
which can be tested  without discussing non--linear evolution. 

More in detail, we shall consider the following prescriptions: 
(i) We set the numerical constant $A_{\Psi}$, in the spectrum (1.1), so that
$C_2$ is consistent with COBE 3$\, \sigma$
intervals for $Q_{rms, PS}$ (for different $n$'s). 

(ii) Once the normalization is fixed at small $k$,
we test the large $k$ behaviour, first of all on the $8\, h^{-1}$Mpc
scale. The mass $M_\lambda$, within a sphere of radius $L= \lambda 
h^{-1}{\rm Mpc}$, is
$$
M_\lambda = 
5.96 \cdot 10^{14} \Omega h^2 M_\odot (\lambda/8)^3~.
\eqno (3.1)
$$
Therefore, the cumulative cluster density
$$
n(>M) = \sqrt{2/\pi} (\rho/M) \int_{\delta_c/\sigma_M}^\infty
du [M/M(u)] \exp(-u^2/2)
\eqno (3.2)
$$
for $M = 4.2h^{-1} \cdot 10^{14}$M$_\odot$ is directly related to the value
of $\sigma_8$. In eq.~(3.2) we take $M(u)$
defined so that the mass variance (evaluated with a top--hat
window function) $\sigma_{M(u)} = \delta_c/u$; $\delta_c$ values from
1.69 (Peebles, 1980) to 1.55 were taken in figures 2 and 3.

Let us then consider $N_{cl} = n(>M) (100h^{-1} {\rm Mpc})^3$
for the above $M$ value. Optical and X--ray observations give a
value of $N_{cl}$ which is still not so different from
$\sim 4$, as found by White at al. (1993) and Biviano et al. (1993).
Both the observational value 
and its theoretical prediction are however subject to a number of
uncertainties. For instance, observations have
some problems to fix cluster masses (see also below).
From the theoretical side, non--linearity
effects and mechanism turning
fluctuations into clusters cannot be said to be completely under control.
It is also wise recalling that $N_{cl}$ and $\sigma_8$ feel the slope of the
spectrum, around  $8 \, h^{-1}$Mpc, in a slightly different way.
Henceforth, we shall further comment on $\sigma_8$ values, after discussing
the spectral slope.

The above arguments tell us that, in our systematic search, 
it is wise to keep models with $1 \mincir N_{cl} \mincir 10$.

(iii) Models can survive the previous test for the whole range of 
$A_{\Psi}$ or for a part of it. Allowed $A_\Psi$ values were then used 
to evaluate the expected density parameters $\Omega_{\rm gas} 
= \alpha \Omega_b \Omega_{\rm coll}$ in damped Lyman $\alpha$ systems 
(for a review  see Wolfe, 1993). Here 
$$
\Omega_{\rm coll} = {\rm erfc}[\delta_c/\sqrt{2} \sigma(M,z)],
\eqno (3.3)
$$
where $\sigma(M,z) $ is the (top hat) mass variance (for mass $M$ at 
redshift $z$) and $\alpha$ is an efficiency parameter which should be 
$\mincir 1$. More specifically, using such expression, we evaluated
$DLAS \equiv \Omega_{\rm gas} \times 10^{3}  /\alpha $,
taking $z = 4.25$, $\delta_c = 1.69$ and $M=5\cdot 10^{9} h^{-1}M_\odot$.

This choice of values, as well as the choice of a {\sl top--hat}
window function, is a compromise among various 
suggestions in the literature, supported
by different arguments. E.g., values of $\delta_c$
greater than 1.7 (Ma $\&$ Bertschinger 1994) or
as small as 1.3 (Klypin et al. 1995) were discussed, 
a value of $M$ ten times greater was often considered and
the quantitative relevance of changing the window function
was debated (see, e.g., Borgani et al., 1996
and Pierpaoli et al., 1996).
According to Storrie--Lombardi et al. (1995), observations give $DLAS
= 2.2 \pm 0.6$. Therefore, we passed models only when $DLAS > 0.5$.
It must be outlined that varying this limit by a factor $\sim 2$ would
cause only marginal changes for models accepted. Slight shifts of
$\Omega_h$ or $d$ usually cause significant variation of $DLAS$
and this constraint turns out to be a fairly substantial one.
Other tests concerning objects at large $z$, like early galaxies
and QSO's, are less restrictive. 

(iv) Bulk velocities were also 
evaluated and compared with POTENT reconstructions of velocity fields.
Here we shall report no detail on the procedure, which causes no
constraint, at the 2$\, \sigma$ level.

On the contrary, a severe further selection arises from requiring that: (v) the
{\sl extra--power} parameter $\Gamma = 7.13 \cdot 10^{-3} (\sigma_{ 8}/\sigma_{
25})^{10/3}$ (here $\sigma_{8,25}$ are mass variances on the scales $R = 8,25\,
h^{-1}$Mpc) has values consistent with observations. From APM galaxies,
Peacock and Dodds (1994) obtained $\Gamma = 0.23 \pm 0.04$. In a more recent
work, Borgani et al. (1997) give the interval 0.18--0.25 obtained from the
Abell/ACO sample. However, owing to the high value (10/3) of the exponent, a
shift of $\sigma_{ 8}$ by a small factor induces a major $\Gamma$ displacement.
For example, if non--linear effects are under--evaluated by just $\sim
10\, \%$, $\Gamma$ encreases by $\sim 37\, \%$. This effect is more
likely for low $\Gamma$ values, which can be due to fairly high
$\sigma_8$ values. Henceforth we kept 0.27 as top acceptable value,
which is both the upper limit given for APM and the $\sim 3\, \sigma$
upper limit for Abell/ACO. On the contrary we took 0.13 as lower limit,
assuming that an underestimate of non--linear effects by $\sim 6$--8$\, \%$
cannot be excluded.

As already mentioned, the constraint on $\Gamma$, together
with a suitable constraint on $\sigma_8$, implies a constraint on
$N_{cl}$. Checks of $\sigma_8$, $N_{cl}$ and $\Gamma$ are, therefore,
strictly related.

For instance, an observational value for $\sigma_8$ 
can be deduced from X--ray data on the gas temperature $T_g$ in galaxy 
clusters. If clusters are substantially virialized and the intracluster 
gas is isothermal, the mass $M$ of a cluster can be obtained, once the ratio
$$
\beta 
= {{\rm galaxy~kinetic~energy/mass} \over {\rm gas~thermal~energy/mass}}
$$
is known; then, the temperature function can be converted
into a mass function, which can be fit to a Press $\&$ Schechter
expression (see eq.~3.1), yielding the normalization of $\sigma(M)$ 
(variance as a function of the mass scale $M$) and hence $\sigma_8$.

An alternative possibility, of course, amounts to deducing masses
from galaxy velocities obtained from optical data.
Both pattern imply some problem.

The value of $\beta$ is currently obtained from numerical models.
Henry $\&$ Arnaud (1991), assuming $\beta = 1.2$, estimated $\sigma_8 = 
0.59 \pm 0.02$ from a complete X--ray flux--limited sample of 25 clusters 
they compiled. Various authors followed analogous patterns (see, e.g., 
White et al. 1993, Viana $\&$ Liddle 1996). Eke et al. (1996), adding 
observational uncertainties and $\beta$ error, claimed that 
$\sigma_8 = 0.50 \pm 0.04$. 
An essential issue to obtain such result is that
Navarro et al. (1995) simulations allow to take $\beta = 1$ and
an error $\mincir 6\, \%$.


However, there seems to be a conflict between expected
galaxy velocities, obtained assuming $\beta = 1 \pm 0.06$ and optical data.
In fact, the latter give a virial velocity dispersion in clusters $\sigma_v 
\simeq 800$  km/s, a value
consistent with $\beta \magcir 1.5$ (Zabludoff et al. 1990, Girardi et al. 
1993). 

It is possible that the critical assumption is that clusters 
are isothermal. In many clusters, cooling flows may
play an important role and a fit to a cooling--flow
cluster with a simple isothermal model may yield a mean emission--weighted 
temperature significantly reduced in respect to the virial value
(see, $e.g.$, Allen $\&$ Fabian 1998). However, according
to Eke et al. (1998), cooling flows
would cause a $z$--trend in disagreement with available data.
More data on high--$z$ clusters are however needed to strengthen
this statement. Other authors (Frenk et al. 1990,
Borgani et al. 1997) claimed that the conflict between predicted and observed
$\sigma_v$ originates from contamination of optical data by
groups accreting onto the clusters.

We can conclude that models with $\sigma_8$ in an interval 0.45--0.75
should be viable. As expected, models passing previous tests keep
comfortably inside such interval. The top $\sigma_8$ needed by models
in fig.~1b is 0.70 (for $\Omega_h = 0.26$, $d=1$, $n=1.2$) 
and the bottom $\sigma_8$
is 0.50 (for $\Omega_h = 0.18$, $d=16$, $n=1.2$). If the constraint
on $\Gamma$ is dropped (fig.~1a), just two models with $\sigma_8 = 0.70$ 
(both yielding $\Gamma < 0.10$) and a model with $ \sigma_8 = 0.48$
($\Omega_h = 0.22$, $d=16$, $n=1.3$, $\Gamma = 0.11$) are added.

Some of the transferred spectra we obtain will be plotted, together
with spectral data obtained from Las Campanas survey (LCRS), kindly
provided by Lin et al. (1996). It is however worth outlining soon that such 
comparison, although suggestive, is not so discriminatory.

A detailed description of LCRS is given by Schectman et al. (1996).
The survey encloses 3+3 ``slices" $1^o.5 \times 80^o$ wide, in the northern
and southern emispheres. Data were taken using two multifiber systems.
Fields $1^o.5 \times 1^o.5$ wide were inspected above suitable
photometric limits, chosen so that there were more galaxies per
field than available fibers. Then, target galaxies in each field
were randomly selected and a ``field sampling factor" $f$ was memorized,
to be used in any further statistical analysis. The average
values of $f$ are different for the two multifiber systems,
which are able to inspect 50 and 112 objects, and amount
to 0.58 and 0.70, respectively. The nominal photometric limits
are also different for the two systems, amounting to $16 \leq m \leq 17.3$ and 
$15 \leq m \leq 17.7$, respectively. A further geometric effect
is due to the impossibility to inspect galaxies, in a given field,
if nearer than 55".

The actual situation for power spectrum measurement from LCRS appears
to fall into two regimes. On scales $L < 80$--$100\, h^{-1}$Mpc
($k > 0.2\, h$) a fair determination of the spectrum is obtained.
In this range, LCRS results strengthen results from
other surveys. Larger scales would be more discriminatory, but here 
errors are greater and the sample variance might cause further shifts.

When compared with such observational spectra, a reasonable
transferred spectrum should lie {\sl below} observational errorbars, up to
$\sim 80$--100$\, h^{-1}$Mpc; the gap between theoretical and observational
spectra is related to the amount of {\sl bias} (constant gap
means a scale--independent bias level). For $k$ values much beyond this scale,
requiring a detailed fit may be excessive. Although there are mixed models
which provide it, we see no reason to disregard models whose
spectrum falls within $\sim 3\, \sigma$ errorbars from the 
reconstructed spectrum. An example of such models are $\Lambda$CDM models, 
with various $\Lambda$ contents, which were shown also by Lin et al. (1996), 
but not treated in this article.

The basic issue following this discussion, which can be
suitably rephrased for other data sets, is that it is non--trivial
to extract discriminatory criteria from observational spectra.
Most of such criteria are expressed by quantities already
introduced earlier in this section. Furthermore, quite in general,
in a systematic search, predictions are to be fit with fairly wide 
observational intervals. In fact, a marginal agreement with the data, for a 
model met in a systematic analysis, can be often improved by a suitable finer
tuning of parameters. We verified that, in several cases, this was the actual 
situation. 

\section{ Inflationary models yielding $n>1$ }

This section is a quick review of results in the literature,
aiming to show that there is a wide class of inflationary
models which predict $n > 1$, but $\mincir 1.4$. It can be therefore skipped
by those who are aware of such results.

Density perturbations arise during inflation because of quantum
fluctuations of the {\sl inflaton} field $\varphi$. Their amplitude
and power spectrum are related to the Hubble parameter $H$ during inflation 
and to the speed $\dot \varphi$ of the {\sl slow--rolling--down} process, 
along the scalar field potential. It can be shown that the critical
quantity is the ratio $W(k) = H^2/\dot \varphi$, where $H$ and $\dot \varphi$
are taken at the time when the scale $2\pi/k$ leaves the event 
horizon. The value of the spectral index can then be shown to be
$$
n = 1 + 2 {d(\log\, W) \over d(\log\, k)}
\eqno (4.1)
$$
and, if $W$ (slowly) decreases with time, we have the standard case
of $n$ (slightly) below unity (it should be reminded that
greater scales flow out of the horizon at earlier times). 
This decrease is due to an acceleration of the downhill motion of
$\varphi$ and an opposite behaviour occurs if $\dot \varphi$
decreases while approaching a minimum. The basic reason why a potential
yielding such a behaviour seems unappealing, is that the very last stages of
inflation should rather see a significant $\varphi$--field acceleration,
ending up into a regime of damped oscillations around the true vacuum, when 
reheating occurs.

These objections can be overcame if the usual perspective is reversed:
reheating does not arise when an initially smooth acceleration finally
grows faster and faster, as the slope of the potential steepens; 
on the contrary, reheating starts abruptly, thanks to a
first order phase transition, perhaps to be identified with the
break of the GUT symmetry. Before such transition and since
the Planck time, most energy content naturally resided in potential
terms, so granting a vacuum--dominated expansion. This picture of
the early stage of the cosmic expansion is the so--called {\sl hybrid 
inflation}, initially proposed by Linde (1991a).

A toy--model realizing such scenario (Linde 1991b, 1994) is obtained from
the potential
$$
V(\varphi,\chi) = (\mu^2 - \lambda \chi^2)^2 + 2g^2 \varphi^2 \chi^2
+ m^2 \varphi^2
\eqno (4.2)
$$
depending on the two scalar fields $\varphi$ and $\chi$, expected to
be slowly and fastly evolving, respectively. If the slowly evolving
field is embedded in mass terms, the potential reads
$$
V(\chi) = M^2 \chi^2 + \lambda \chi^4 + \Lambda^4
\eqno (4.3)
$$
where
$$
\Lambda^4 = \mu^4 + m^2 \varphi^2 ~~~~~{\rm and} ~~~~~~
M^2 = 2(g^2 \varphi^2 - \lambda \mu^2) ~.
\eqno (4.4)
$$
Eq.~(4.3) shows that $V$ has a minimum at $\chi = 0 $, provided that $M^2 > 0$.
If  $M^2 < 0$, instead, the minimum is for $\bar \chi$$= 
\sqrt{-M^2/2\lambda}$, yielding $\mu$ when $\varphi = 0$.

Large $\varphi$ values therefore require that $\chi$ vanishes and then
the potential
$$
V(\varphi,0) = \mu^4 + m^2 \varphi^2
\eqno (4.5)
$$
allows a Planck--time inflation, as $\varphi$ rolls downhill taking
$V$ from an initial value $\sim t_{Pl}^{-4}$ to $\sim \mu^4$. This downhill
motion is expected to decelerate when the second term at the $r.h.s.$
of eq.~(4.5) becomes negligible in respect to $\mu^4$, which essentially
acts as a cosmological constant. This deceleration abruptly breaks down
when the critical value $\varphi_c = \sqrt{\lambda} \mu/g$ is attained,
for which $M^2$ changes sign. At that point the configuration $\chi = 0$
is unstable and the transition to the true vacuum configuration $\bar \chi$
reheats (or heats) the Universe.

There are a number of constraints to the above picture, due to the
requirements that at least 60 e--foldings occur with $\varphi > \varphi_c$
and that fluctuations have a fair amplitude. Such constraints are
discussed in several papers (see, $e.g.$, Copeland et al. 1994, and 
references therein) and cause the restriction $n \mincir 1.4$. Henceforth,
in this work we shall debate models with $1 \leq n \leq 1.4$.

Let us however outline that {\sl hybrid} inflation is not just one
of the many possible variations on the inflationary theme. In spite of
the apparent complication of the above scheme, it is an intrinsically simple
picture and one of the few
patterns which can allow to recover a joint particle--astrophysical
picture of the very early Universe, as was naturally hoped before
it became clear that the Higgs' field of the GUT transition could not
be the inflaton (see, $e.g.$, Shafi $\&$ Vilenkin, 1984 and Pi, 1984).

\section{ Transfer functions and spectra}

Any realistic cosmological model is expected to contain baryons (density 
$\rho_b = \Omega_b \rho_{cr}$), photons (present temperature $T_o = 2.733 
\theta \, $K, density $\rho_\gamma$), CDM (density $\rho_c = \Omega_c 
\rho_{cr}$) and a SMLC. The basic SMLC are the usual 3 massive $\nu$'s.

Mixed models include also a HDM component
(density $\rho_h = \Omega_h \rho_{cr}$) and the SMLC is however modified.
If HDM particles are massive $\nu$'s, SMLC density is lower, or even 
vanishing 
when all $\nu$'s are massive (of course, sterile massless particles can be
however added {\sl ad hoc}). In the volatile case,
we face the opposite situation, as the SMLC must contain at least a scalar,
in top of usual massless $\nu$'s. In all cases, we can parametrize SMLC by
using $w =  0.68132(g_{\nu,eff}/6)$, according to eq.~(2.14).

The algorithm used to evaluate the transfer functions of a wide range of
mixed models is discussed  in Appendix B and, besides of allowing any 
value of $\Omega_b$, $\Omega_h$, $w$ and other 
standard parameters, allows to fix arbitrarily the expression
of $\Phi (p)$ [see eqs.~(2.1) and (2.7)] and the dependence
on $p$ of the amplitude of HDM fluctuations.

As already mentioned, however, in this work $\Omega = 1$, $\Lambda = 0$,
$H=50\, $km$\, $s$^{-1}$Mpc$^{-1}$ and $\Omega_b = 0.1$
(some plots for thermal models, with different values of
$\Omega_b$ are however shown). Exploring other
values of the former 2 parameters would amount to extending the
physical range of models. The expected behaviour for different values of
the latter two parameters, instead, can be qualitatively inferred
from the results shown below. A test of 
the effect of varying them should be focused onto a portion of the 
parametric space, selected on the basis of the results of this work
and other physical criteria.

Within the above restrictions, we tested models at regular logarithmic
interval for $z_{der}$, taking $\log_2 d = -1,...,5$ (7 values) and at regular
interval for hot--dark--matter density, taking $\Omega_h = 0.10,\,
 0.12,\, ....,\, 0.44, {\rm ~and~ }0.45$ (19 values). 
A non--systematic sampling of this
parameter space allowed us to exclude some models a priori, reducing
the total exploration to 120 models. 

There is a significant overlap between these values of $\Omega_h$ and $d$ 
and those allowed in mixed models with massive $\nu$'s, as can be seen also 
in fig.~1 herebelow. One of our basic aims was then to obtain a
quantitative estimate of the impact of the different momentum
distribution on transfer functions. To this aim we re-evaluated
$\cal T$, for a few mixed models with HDM made either by
massive neutrinos or volatiles, but with the same values of
$\Omega_h$, $d$ and $g_{\nu,eff}$, according to eqs.~(2.3),~(2.4),~(2.14).
The results of this comparison for two model sets are shown in Table I.


\begin{table}
\caption{ Ratio between tranfer functions in volatile and thermal models
(${\cal T}$ (vol)/ ${\cal T}$ (ther))
 at various comoving scales for different components.
The upper part refers to models with $\Omega_h = 0.3$ 
 and 3 massive neutrinos, the lower part to models with
$\Omega_h = 0.1$ and 1 massive neutrino.
In both cases $\Omega=1$, $h = 0.5$ and $\Omega_b = 0.1$.}
\begin{tabular}{rrrrr} 
\hline
L/Mpc & cold  &  hot & baryons & total\\
\hline
    0.2000E+05  & 1 & 1 & 1 & 1\\     
    0.5120E+03 & 1 & 1 & 1 & 1\\    
    0.2048E+03  & 0.9995E+00 & 0.9996E+00 & 0.9995E+00 & 0.9995E+00\\
    0.8192E+02  & 0.1001E+01 & 0.1002E+01 & 0.1001E+01 & 0.1001E+01\\
    0.3277E+02  & 0.1003E+01 & 0.1004E+01 & 0.1003E+01 & 0.1003E+01\\
    0.1311E+02  & 0.1003E+01 & 0.1004E+01 & 0.1003E+01 & 0.1003E+01\\
    0.5243E+01  & 0.1005E+01 & 0.1003E+01 & 0.1005E+01 & 0.1004E+01\\
    0.2097E+01  & 0.1007E+01 & 0.1012E+01 & 0.1007E+01 & 0.1009E+01\\
    0.8389E+00  & 0.1006E+01 & 0.1036E+01 & 0.1006E+01 & 0.1015E+01\\
\hline
\hline
    0.2000E+05 &     1        &    1      &    1        &    1     \\
    0.5120E+03 &     1        &    1      &    1        &    1     \\
    0.2048E+03 &   0.9995E+00 & 0.9996E+00 & 0.9995E+00 & 0.9995E+00\\
    0.8192E+02 &   0.1001E+01 & 0.1001E+01 & 0.1001E+01 & 0.1001E+01\\
    0.3277E+02 &   0.1003E+01 & 0.1002E+01 & 0.1003E+01 & 0.1003E+01\\
    0.1311E+02 &   0.1003E+01 & 0.1002E+01 & 0.1003E+01 & 0.1003E+01\\
    0.5243E+01 &   0.1005E+01 & 0.1001E+01 & 0.1005E+01 & 0.1005E+01\\
    0.2097E+01 &   0.1007E+01 & 0.1008E+01 & 0.1007E+01 & 0.1007E+01\\
    0.8389E+00 &   0.1006E+01 & 0.1026E+01 & 0.1006E+01 & 0.1008E+01\\
\hline
\end{tabular}
\end{table}

The detection of the (modest) amount of such shifts is one 
of the results of this work. However, this does not decrease the relevance
of volatile models, as they allow parameter choices which, otherwise,
would be impossible with the standard 3 $\nu$--flavours. Furthermore,
these models require and allow a high level of SMLC.

The main output of this work is however represented by fig.~1(a,b),
where we show which parts of the parameter space are compatible with
linear constraints, for spectral indices $n = 1,1.1,...,1.4$.

Fig.~1b takes into account all the constraints mentioned in sec.~3.
Models inside the curves fulfill the requirements on COBE quadrupole
value (3$\, \sigma$ intervals), galaxy clusters, high--z objects
and $\Gamma$. 
On the contrary, fig.~1a excludes the requirement on $\Gamma$.
Models forbidden by fig.~1b but
allowed by fig.~1a should be accurately  tested for $\Gamma$ in
non--linear simulations.

For most models considered, the spectrum
$$
P (k) = A k^n {\cal T}^2(k),
\eqno (5.1)
$$
where the constant
$$
A={2 \pi^3 \over 9} x_o^{n+3} A_{\Psi},
\eqno (5.2)
$$
starts -- at low $k$'s -- with values similar to 
standard CDM. If $n>1$, it soon abandons CDM behaviour,
raising in a slightly steeper way. Both for $n>1$ and $n=1$, however,
its bending at maximum is sharper. At the $r.h.s.$ of the maximum
it returns  below CDM. At large $k$'s ($\sim 1\, $Mpc$^{-1}$), its
decrease, sometimes, is (slightly) less steep than CDM.

Some examples of such behaviour are given in fig.~2 for volatile
models and in fig.~3 for thermal models; all
cases we show have $n>1$. Models are compared with CDM (dotted curve)
and with the spectrum reconstructed by Lin et al (1996)
from LCRS data (3$\, \sigma$ errorbars).
Models are ordered with increasing $\Omega_h$. For volatile models,
various values of $d$ are considered, including high values; in particular,
in fig.~2c we show a spectrum obtained with $z_{der} = 625$.

Massive neutrino models obviously have $w < w_o$.
Extra SMLC can be however added {\sl ad hoc} and
the comparison of Table I was made after implementing $w$ to the value
of volatile models. The boundaries shown in fig.~1 are also valid
when all SMLC expected in a volatile model is included.
Its presence causes a later transition from radiation dominated
to matter dominated expansion. As a consequence, the peak of the transferred
spectrum is shifted to smaller $k$'s and, in general, this favours
the agreement of models with linear constraints.

While the boundaries marked in fig.~1 therefore apply to neutrino
models with extra SMLC, the spectra shown in fig.~3(a--b) are for physical
models without extra SMLC. The spectrum in fig.~3c, instead,
is obtained adding an amount of SMLC corresponding to 3 massless $\nu$'s.

The inverse transfer functions $\cal T$$^{-1}$ of all models considered,
were also parametrized with 4--th degree polynomials in $\sqrt{k}$,
both for $z=0$ and $z=4.25$. The table of coefficients will be made 
available upon request.

\section{ Discussion}

After evaluating the transfer function of a set of more than 120
mixed models, with $\Omega = 1$ and $h=0.5$, we obtained the
expected values of several observable quantities, which can be
estimated using the linear theory. In particular we worked out
the expected cluster number density, the abundance of Damped Lyman $\alpha$
clouds, the mass variances at 25 and 8$\, h^{-1}$Mpc. From them we
also estimated the {\sl extra power} parameter $\Gamma$.

The HDM of the models we considered was either thermal (massive $\nu$'s)
or volatile (arising from heavier particle decay). We discussed the
parameter space for the latter case and compared the transfer functions
for thermal and volatile HDM, for values of $\Omega_h$ and $d$ which
allow both models. Differences between thermal and volatile models are then
really significant only as far as the expected SMLC is concerned.

Results on parameter constraints are summarized in figures 1 (a,b).
As was already known, a number of such models with $\Omega_h$
up to 0.30 pass the above tests. If $n=1$,
volatile models allow little extra freedom, namely for high $z_{der}$.
The situation is already different
for $n \simeq 1.1$. Here models with $z_{der} \simeq 600$ are allowed
for $\Omega_h $ up to 0.14 and greater $\Omega_h$ are allowed for
values of $z_{der}$ still much lower that those allowed by neutrino models.
The range of $\Omega_h$ values allowed with $z_{der} \simeq 600$ extends
upwards as $n$ increases and overcomes 0.20 for $n = 1.4$.
The greatest value found for $\Omega_h$ is 0.30 for $z_{der} = 10^4$
with $n=1.2$. Thermal models with 2 or 3 massive neutrinos
and a suitably added SMLC arrive to $\Omega_h \simeq 0.28$, for $n=1.2$
and $n=1.3$, respectively.

These values may not seem too large, in respect to $\Omega_h$ values
currently used in the literature (with thermal models). It must be
outlined, however, that our {\sl acceptance }criteria are more
stringent than usual. With such criteria, no thermal model with
$\Omega_h > 0.16$ is {\sl accepted} for $n \leq 1$.

Let us also draw the attention on the very low values of $z_{der}$ that
volatile models with large $n$ allow.
As an example of low $z_{der}$, in fig.~2c we show the transferred spectrum of a
model with 22$\, \%$ of HDM and $z_{der}= 625$, which, for a fairly
high value of COBE quadrupole, has excellent fits with all linear constraints.
HDM particles of this and similar models have a mean square velocity
$\sim 500\, $km/s today, still away from potential wells. Such speed
is likely to guarantee them not to cluster with baryon or CDM
on any scale. Dynamical mass estimates, in a world containing such
component, might lead to observe $\sim 75$--$80 \, \%$ 
of critical density.

It should be also mentioned that, if the constraint on $\Gamma$ is dropped, 
greater $\Omega_h$'s seem allowed (see fig. 1a). The top value we found is
$\Omega_h = 0.45$ with $z_{der} \simeq 5000$. With 3 or 2 massive neutrinos
the highest $\Omega_h$ obtainable are $0.40$ and $0.38$, respectively.
As already outlined, in fig.~1 neutrino model
curves are overlapped to volatile model parameter space, but
the allowed regions apply to them only after adding suitable SMLC.

Liddle et al. (1996), in a paper focused on $n \leq 1$ models, mentioned
that a model with $\Omega_h = 0.35$, $n=1.2$, $h=0.4$ agreed with the
linear constraints they imposed. Using our linear constraints,
$\Omega_h$ is to be lowered to $\sim 0.30$.
Volatile models do not need to lower $h$ so far, to agree with
data, as they naturally have SMLC, which induces fairly similar effects.
Also thermal models can agree with data
keeping to $h=0.5$, rising $n$, with or without extra SMLC. Fig.~3
illustrates some spectra for such models.

Models with high HDM content and high average kinetic energy,
could also ease the problem of the apparent baryon excess in
several galaxy clusters.
It has been known for several years (see, e.g., White $\&$ Frenk 1991,
Boehringer et al. 1992, 
Briel et al. 1992, David et al. 1993, White et al. 1993) that the baryon mass 
fraction in galaxy clusters is high, in respect to expectations in flat, 
pure CDM models. According to White et al. (1993),
even neglecting baryons in galaxies, the {\sl observed} baryon fraction yields
$$
\Omega_b h^{3/2} = 0.05 \pm 0.02
\eqno (6.1)
$$
for flat, pure CDM models. BBNS limits of Walker et al (1991)
required then that $\Omega_b h^2 = 0.0125 \pm 0.0025$.

Much work was then performed to analyze individual clusters
and lists of baryon contents in cluster samples were presented by
several authors (see, e.g., White $\&$ Fabian, 1995).
The situation is summarized in a recent paper by Evrard (1997)
who obtains the constraint
$$
\Omega_b h^{3/2} = 0.060 \pm 0.003
\eqno (6.2)
$$
This is to be taken with recent BBNS limits on 
$\Omega_b$. $E.g.$, Copi et al (1995) give $0.007 < \Omega_b h^2 < 0.024$
and therefore $\Omega_b \simeq 0.1$, with $h=0.5$ is marginally allowed.
For $h = 0.5$, as we assumed here, eq.~(6.2) gives a central value
$\Omega_b = 0.17$ and a 3$\, \sigma$ minimum $\Omega_b \simeq 0.14$.

If HDM does not cluster with CDM and baryons,
a measure of $\Omega_b$ would however yield $\Omega_{b,app} =
\Omega_b/(1-\Omega_h)$. 
Within mixed models, therefore, agreement with data requires
$\Omega_h > 0.41$ or $> 0.29$, respectively. We have shown in
this work that the latter value is coherent
with LSS constraints, if $1.2 \leq n \leq 1.4$; as $n$ increases, compatible
models have $z_{der}$ values decreasing, down to $\sim 2500$.

Limits become more permissive if
masses obtained from lensing are replaced to those obtained from
X--ray data (see, e.g., Allen $\&$ Fabian 1998).
High baryon density, however, seems a widespread
feature. Perhaps the limits shown in eq.~(6.2)
are to be reset, taking greater cluster masses, but,
if $\Omega_o=1$, there seems to be a definite evidence of
a component which tends not to cluster with CDM.
The debate is related to the discussion on $\sigma_8$ limits in sec.~3.

In the literature, mixed models with $n \leq 1$ were often
considered and allow to predict acceptable values of a number of
observable quantities, if low rates for the HDM
component are taken. Considering $n > 1$ is at least as legitimate as taking
$n < 1$ and leads to a range of mixed models allowing
fair predictions on the same quantities.
We showed that models with $n > 1$ require and allow higher HDM contents.
They could therefore improve our understanding of why $\Omega_o$ measures
seem to give increasing values when greater scales $L$ are considered.
This was one of the basic motivations to introduce mixed models, 
more than a decade ago.

\ack{Thanks are due to Stefano Borgani for discussions. E.P. wishes to 
thank the Department of Physics of the University of 
Milan for its hospitality during part of the preparation of this work.
E. Bertschinger, who refereed this paper, is to be gratefully thanked
for a number of useful suggestions. Thanks are also due to H. Lin,
who kindly gave us points and errorbars for the LCRS spectrum. }

\vfill\eject

\parindent=0.truecm
\parskip 0.1truecm

\vskip 0.2truecm
{\bf References}

\vskip 0.1truecm

Achilli S., Occhionero F. and Scaramella R., (1985), {\APJ} 299, 577

Allen S.W. and Fabian A.C. (1998) astro-ph/9802219 

Atrio--Barandela F., Einasto J., Gottoeber S., Mueller V. and
Starobinski A. (1997) (preprint)

Bennet C.L. et al. (COBE collaboration) (1996), {\APJ} 464, L1 

Biviano A., Girardi M., Giuricin G., Mardirossian F. and Mezzetti M. (1993)
{\APJ} 411, L13

Bonometto S.A., Gabbiani F. and Masiero A. (1994)  {\PRD} 49, 3918

Bonometto S.A. and Valdarnini R. (1984) {\PLA} 103, 369

Borgani S., Gardini A., Girardi M. and Gottloeber S. (1996) {\MN}
280, 749

Borgani S., Lucchin F., Matarrese S. and Moscardini L. (1996) {\MN}
280, 749

Borgani S., Moscardini L., Plionis M., G\'orski K.M., Holzman J.,
Klypin A., Primack J.R., Smith C.C. and Stompor R. (1997)
{\em New Astr.} 321, 1

Broadhurst T.J., Ellis R.S., Koo D.C. and Szalay A. (1990) {\Na} 343, 729

Burigana C., De Zotti G. and Danese L. (1991) {\APJ} 379, 1

Carr B.J. and Lidsey J. (1993) {\PRD} 48, 543

Colombi S., Dodelson S. and Widrow L.M. (1996) {\APJ} 458, 1

Copeland E.J., Liddle A.R., Lyth D.H., Stewart E.D. and Wands D. (1994) 
{\PRD} 49, 6410

Copi C.J., Schramm D.N. and Turner M.S. (1995) {\SCI} 267, 192

Dodelson S., Gates E. and Stebbins A., (1996) {\APJ} 467, 10

Einasto J., Einasto M., Gottloeber S., Mueller V., Saar V., Starobinski
A., Tago E., Tucker D., Andernach H. and Frisch (1997) {\Na} 385, 139

Eke V.R., Cole S. and Frenk C.S. (1996) {\MN} 282, 263

Eke V.R., Cole S., Frenk C.S. and Henry J.P. (1998) astro-ph/9802350

Evrard A.E. (1997) {\MN}  292, 289

Frenk C.S., White S.D.M., Efstathiou G. and Davis M. (1990) 
{\APJ} 351, 10

Gaztanaga E. and Baugh C.M. (1997) astro--ph/9704246

Ghizzardi S., and Bonometto S.A., (1996) {\AA} 307, 697

Girardi M., Biviano A., Giuricin G., Mardirossian F. and Mezzetti M.
(1993) {\APJ} 404, 38

Hancock S., Rocha G., Lasenby A.N. and Gutierrez C.M. (1998), {\MN}
294, L1 

Henry J.P. and Arnaud K.A. (1991) {\APJ} 372, 410

Klypin A., Borgani S., Holtzmann J. and Primack J.R. (1995) 
{\APJ} 444, 1
 
Landy S.D., Shectman S.A., Lin H., Kirshner R.P., Oemler A.A. and
Tucker D. (1996) {\APJ} 456, L1

Liddle A.R., Lyth D.H., Schaefer R.K., Shafi Q. and Viana P.T.P.
(1996) {\MN} 281, 531

Lin H., Kirshner R.P., Shectman S.A., Landy S.D., Oemler A., 
Tucker D.L. and Schechter P. (1996)  {\APJ}  471, 617 

Lineweaver C. (1997) astro--ph/9702040

Lineweaver C. and Barbosa D. (1998) {\APJ} 496 (in press).

Linde A. (1991a) {\PLB} 249, 18

Linde A. (1991b)  {\PLB} 259, 38

Linde A. (1994) {\PRD} 49, 748

Lucchin F., Colafrancesco S., De Gasperis G., Matarrese S., Mei S.,
Mollerach S., Moscardini L. and Vittorio N. (1996) {\APJ} 459, 455

Ma C.P. and Bertschinger E. (1994)  {\APJ} 434, L5

Ma C.P. and Bertschinger E., (1995) {\APJ} 455, 7

Mollerach S., Matarrese S. and Lucchin F. (1994) {\PRD} 50, 4835

Netterfield C.B., Devlin M.J., Jarosik N., Page L. and Wollack E.J. 
(1997) {\APJ} 474, 47

Olive K., Steigman G. and Skillman E., (1997) {\APJ} 483, 788

Peacock J.A. and Dodds S.J. (1994) {\MN} 267, 1020

Peebles P.J.E. (1980), The Large Scale Structure of the Universe, 
Princeton University Press, Princeton.
 
Pi S.Y. (1984) {\PRL} 52, 1725

Pierpaoli E. and Bonometto S.A. (1995) {\AA} 300, 13

Pierpaoli E., Coles P., Bonometto S.A. and Borgani S. (1996) {\APJ} 470, 92

Pierpaoli E., Borgani S., Masiero A. and Yamaguchi M., (1998)
{\PRD} 57, 2089

Seljak U. and Zaldarriaga M. (1996) {\APJ} 469, 437

Shafi Q. and Vilenkin A. (1984) {\PRL} 52, 691

Shectman A.S., Landy S.D., Oemler A., Tucker D.L., Kin H.,
Kirshner R.P. and Schechter P.L. (1996) {\APJ} 470, 172 

Storrie--Lambardi L.J., McMahon R.G., Irwin M.J. and Hazard C. (1995) {\em
Proc. ESO Workshop on QSO A.L.} $\&$ astro--ph/9503089

Valdarnini R. and Bonometto S.A., (1985),{\AA} 146, 235

Viana P.T.P. and Liddle A.R. (1996) {\MN} 281, 323

Walker T.P., Steigman G., Schramm D.N., Olive K.A. and Kang H.
(1991) {\APJ} 376, 51

White S.D.M., Efstathiou G. and Frenk C. (1993) {\MN} 262,1023

White S.D.M. and Frenk C. (1993) {\APJ} 379,52

Wolfe A., in {\em Relativistic Astrophysics and Particle Cosmology}, ed.
Ackerolf C.W. and Srednicki M.A. (New York Acad. Sci., New York, 1993)

Zabludoff A.I., Huchra J.P. and Geller M.J. (1990) {\APJS} 74, 1

\vfill\eject

\centerline{ FIGURE CAPTIONS }

\vglue 0.1truecm
Fig.~1 -- The curves enclose areas where models predict LSS linear features
consistent with observations. Each curve refers to the value of $n$
written aside. Point--dashed lines refer to models with HDM
made of massive $\nu$'s. See the text for a more detailed comparison
of the LSS they predict with volatile models.
In fig. 1a the constraints taken into account are:
cluster abundance, $\sigma_8$ value, gas in damped Ly--$\alpha$ systems.
In fig. 1b the value of the extra--power parameter $\Gamma$ is also 
constrained.

Fig.~2 --  Mixed model spectra with volatile hot component (solid line), 
compared with standard CDM (dotted line). 
The LCRS reconstructed spectrum is also reported (3$\, \sigma$ errorbars).
Models $a,b,c$ are consistent with all constraints. Model $d$ is outside the
contours plotted in fig.~1b and has a small
$\Gamma$. Models like it would agree with data for a still greater $n$.

Fig.~3 --  Mixed model spectra, with thermal hot component (solid 
line) compared with standard
CDM (dotted line) and LCRS reconstructed spectrum, as in fig.~2.
Models $a,b$ agree with all constraints. Model $c$ is in marginal 
disagreement, in spite of adding extra SMLC amounting to 3 massless $\nu$'s.

\vfill\eject
  
\parindent = 0.5truecm
\noindent
{\bf Appendix A}

\vglue 0.4truecm
\noindent
In the decay $X \to v + \phi$, both $v$ and $\phi$ acquire momentum
(and energy) $p_{in} = m_X/2$. If a decay occurs at $t_p$, at any
later $t$, the momentum is redshifted into $p = p_{in} a(t_p)/a(t)$;
in turn this means that, at any $t$, a $v$--particle of momentum
$p$ was born when the scale factor was 
$$
a(t_p) = a(t)\, 2p/m_X ~.
\eqno (a1)
$$
The {\sl map} of $p$ on $a(t_p)$ yields that $2\, a(t)/m_X = a(t_p)/p$
and
$$
dp/d[a(t_p)] = m_X/2a(t) = p/a(t_p) ~.
\eqno (a2)
$$
This allows to give the momentum distribution of $v$--particles
$$
\Phi_v (p) \equiv {1 \over N_v} { dN_v \over dp }
= { a(t_p)/{\dot a(t_p)} \over p\, \tau_{dy} } 
\exp\left( -{t_p \over \tau_{dy}} \right)
\eqno (a3)
$$
when the decay is {\sl complete}. In fact, using (a2),
$dN_v/dp =$$ [dN_v/dt_p]$ $[dt_p/da(t_p)]$ $[da(t_p)/dp] =$$
 [N_v/\tau_{dy}]$$
[1/{\dot a(t_p)}][a(t_p)/p]$.

The expression (a3) is greatly simplified is the expansion is
radiation dominated during the decay. In such a case $ a(t_p)/{\dot a(t_p)}
= 2t_p$, while, using eq.~(a1), 
$$
Q \equiv t_p/\tau_{dy} = [a(t_p)/a_{dy}]^2 = (p /\tilde p)^2
\eqno (a4)
$$
with
$$
{\tilde p} = (m_X/2)[a_{dy}/a(t)] ~.
\eqno (a5)
$$
Accordingly, 
$$
\Phi_v (p) = 2(Q/p)\exp(-Q)~.
\eqno (a6)
$$
Henceforth, at large $p$, the distribution decays $\propto \exp(-p^2)$.
Large $p$ come from late decays, when the decay process is however radiation
dominated; therefore the latter property is always true.

Owing to eq.~(a4)
$$
dQ/Q = 2\, dp/p
\eqno (a7)
$$
and, therefore, $\Phi_v$ can be easily verified to be normalized to unity.
In a similar way
$$
\langle p \rangle \equiv \int_0^\infty dp \, p\, \Phi_v (p) =
2 {\tilde p} \int_0^\infty dx\, x^2 \exp(-x^2) = {\tilde p} \sqrt{\pi}/2
\eqno (a8)
$$

This allows to replace, in the definition (a5) of $\tilde p$ and, henceforth,
in the distribution $\Phi_v$, parameters referring to the decay process
($m_X$ and $\tau_{dy}$) with parameters explicitly related to a
cosmological model: the derelativization time/redshift/scale factor
(subscript $_{der}$) and the $v$--particle mass $m$. By definition,
let the derelativization occur when $\langle p \rangle = m$ or,
according to eq.~(a8), when $\tilde p$$ = 2m/\sqrt{\pi}$. Then, at
any time:
$$
{\tilde p} = (2m/\sqrt{\pi}) (a_{der}/a) ~.
\eqno (a5')
$$
Henceforth, the distribution peak and cut--off move towards
smaller energies as the Universe expands, with $\tilde p$$a = {\rm const.}$.

In Appendix B, we discuss the time evolution of a model.
The basic reference redshift will be $z_{eq} = 10^4$.
If $\theta = T_o/2.7333\, $K$=1$, the present radiation density
$\rho_{r,o} \equiv \rho_{cr}/z_{eq}$.

In respect to $z_{eq}$ let us define: $d = z_{eq}/z_{der}$,
$y = a_{eq}/a = (1+z) \theta^4/10^4 4h.$ It will be:
$$
\langle p \rangle = myd
~~~{\rm and}
~~~
{\tilde p} = myd\, 2/\sqrt{\pi} ~.
\eqno (a9)
$$

In the case of HDM made of massive $\nu$'s, the momentum distribution reads:
$$
\Phi_\nu (p) = [2/3\zeta(3)] [p/T_\nu]^3 [p(e^{p/T_\nu}+1)]^{-1}
\eqno (n6)
$$
and, accordingly,
$$
\langle p \rangle = [7 \pi^4 / 180 \zeta(3)] T_\nu
\eqno (n8)
$$
Eq.~(n6) shows that, for large $p$, such distribution decays $\propto
\exp(-p)$.

Furthermore, $\nu$'s become non--relativistic when $T_\nu = m
180 \zeta(3)/7 \pi^4$ and, while the former equation (a9) still holds,
it is 
$$
T_\nu = m y d \, 180 \zeta(3)/7\pi^4 ~.
\eqno (n9)
$$

Let us now show that
$$
\rho_h/\rho_\gamma = \Omega_h d ~~~~{\rm at~high~}z.
\eqno (a10)
$$
At such $z$'s, in fact, $\langle \epsilon \rangle = \langle p \rangle$
can be expressed through eq.~(a9). In turn, the $v$ number density
can be obtained rescaling its value at $z=0$. Henceforth
$$
\rho_h =  myd \, n_h = 
myd {\Omega_h \rho_{cr} \over m} \left(a_0 \over a\right)^3 
= \Omega_h d \left[\rho_{cr} \left(a_{eq} \over a_0 \right)\right]
\left(a_0 \over a\right)^4
\eqno (a11)
$$
and, owing to the definition of $a_{eq}$, this shows (a10).

Let us now consider the constraints to a volatile model when $\phi
\equiv \gamma$. The photon density before decay ($\rho_{\gamma,early} $)
has safe lower limits related to BBNS, that we shall not discuss here; 
a fresh $\gamma$ component, whose density coincides with $\rho_h$, accretes
onto it because of the decay; hence, at any time, $\rho_\gamma >\rho_h$
and, owing to eq.~(a10),
$$
\Omega_h d \ll 1 ~~~~~~{\rm for~}\gamma{~decay}.
\eqno (a12)
$$
Under such condition, it is almost granted that no matter dominance
occurs, during the decay process. The actual constraint for this not
to happen is that
$$
\Omega_h d \ll [1+2/(1+g_\nu/8.8)]^{-1} ~~~~~~{\rm for~}\gamma{~decay}.
\eqno (a13)
$$
In fact, comparing densities before and after $\tau_{dy}$, we have 
that the top value of 
$\rho_X \simeq 2\rho_h$ and $\rho_{\gamma.early} \simeq \rho_\gamma-
\rho_h$; furthermore $\rho_\nu = w_o \rho_{\gamma,early}$ with
$w_o = (4/11)^{4/3}7g_{\nu,o}/16 \simeq g_{\nu,o}/8.8$ 
($g_{\nu,o}$ are the spin states
of massless $\nu$'s). The requirement that
the top value of $\rho_X < \rho_{\gamma,early}
+ \rho_\nu$, yields $2\rho_h < (1+w_o) (\rho_\gamma - \rho_h)$ and hence,
owing to eq.~(a10), $\Omega_h d \ll [1 + 2/(1+w_o)]^{-1} $.
$E.g.$, for $g_{\nu,o} = 6$, eq.~(a13) yields $\Omega_h d < 0.46$.

Let us now consider the case of massless sterile scalar $\phi$'s.
Arguments similar to those leading to eq.~(a13) assure no matter
dominance before $\tau_{dy}$ if
$$
\Omega_h d < (1+w_o)/2 ~~~~~~~{\rm (for~sterile~decay).}
\eqno (a14)
$$
In fact, in this case, the
cosmological effects of $\phi$'s add to those of massless $\nu$'s and
we shall therefore distinguish the real $\nu$ density $\rho_{\nu,o}$
from the effective $\nu$ density
$$
\rho_\nu = \rho_{\nu,o} + \rho_\phi = w \rho_\gamma
\eqno (a15)
$$
where $w = w_o + \Omega_h d$. Once again, the top value of $\rho_X$
is $\simeq 2 \rho_h$ and, in order that, at all times,
$\rho_X < \rho_\gamma (1+w_o)$, owing to eq.~(a10), eq.~(a14) 
must be fulfilled. The increase of the radiative
sterile component can be also expressed renormalizing $g_{\nu,o}$
into
$$
g_{\nu,eff} = g_{\nu,o} + 8.80640\, \Omega_h d 
\eqno (a16)
$$
which coincides with eq.~(2.14).

Let us then discuss where the actual equivalence redshift $\tilde z_{eq}$
is set. If $v$--particles do not contribute to the relativistic component,
$$
{\tilde z_{eq}} = \rho_{cr}/(1+w)\rho_\gamma = z_{eq}/(1+w)\theta^4
\eqno (a17)
$$
If $z_{der} < \tilde z_{eq}$, as is defined here, $i.e.$ if $d(1-\Omega_h)
> 1+w_o$, $v$'s are to be added to the relativistic component and
$$
{\tilde z_{eq}} = z_{eq} (1-\Omega_h)/(1+w_o+2\Omega_h d)\theta^4 ~.
\eqno (a18)
$$

\vskip 0.5truecm
\noindent
{\bf Appendix B}

\vglue 0.4truecm
\noindent
The system of equations to be integrated in order to work out
fluctuation evolution has been discussed in many papers. $E.g.$,
an excellent and updated discussion is given by Ma $\&$ Bertschinger (1996).
Here we shall report only the changes needed to treat HDM, if it is made 
of $v$--particles.

Let the phase--space distribution be $f = f_o (1+\epsilon)$.
For massive $\nu$'s
$$
f_o = (2\pi)^{-3} g_\nu (e^x+1)^{-1} ~~~~~{\rm with} ~~~~
x = p/T_\nu
\eqno (n1)
$$
and, therefore, the expression
$$
(x/f_o ) (\partial f_o/\partial x) = -x/(1+e^{-x})
\eqno (n2)
$$
will enter the equation (b3) here below. For volatiles, instead,
$$
f_o = (2 \pi)^{-1} N_v (a_o/ap)^3 Qe^{-Q}
\eqno (b1)
$$
and 
$$
(ap/f_o) (\partial f_o /\partial ap)
= -(1+2Q)
\eqno (b2)
$$
will be used in the equation:
$$
{\partial \epsilon \over \partial t}
+ i{p \over E} k {a_o \over a} \mu \epsilon + {1 \over 4}
H {ap \over f_o} {\partial f_o \over \partial ap} = 0
\eqno (b3)
$$
where $k$ is the wave--number of the perturbation, $H$ accounts
for the gravitational interaction and $\mu$ yields the angle
dependence of $\epsilon$.  It should also be outlined that,
while for massive $\nu$'s, $p/E = [1 + 1/x^2 y^2 d_\nu^2]^{-1/2}$,
with $d_\nu = d\, 180 \zeta(3)/7 \pi^4$, here
$$
{p \over E} = \left[ 1 + { 1 \over Q y^2 d_v^2} \right]^{-1/2}
\eqno (b4)
$$
with $d_v = d\, 2/\sqrt{\pi}$.

The different phase--space distribution has further effects
on the gravitational field equations, that we treated in the synchronous
gauge. Here we consider the stress--energy tensor $T_{ij}$ for massive $\nu$'s
and its perturbation
$$
\Delta(2T_{00} - T) = \int d^3 p E(1-m^2/2E^2) f_o \sigma_o,
\eqno (n4)
$$
where $\sigma_o$ is the 0--th order term in a spherical
harmonic expansion of $\epsilon$. As
$E(1-m^2/2E^2) = m(1/2+x^2 y^2 d_\nu^2)/(1+x^2 y^2 d_\nu^2)^{1/2}$
it turns out that
$$
\Delta(2T_{00} - T) = {1 \over 2} \rho_{cr} \Omega_h z_{eq}^3 y^3
e_{3,\nu} (yd_\nu)
\eqno (n5)
$$
{\rm with}~~~
$$
e_{3,\nu} (s) = 2 \int_0^\infty dx e^{-x} \pi(x) {1/2 + s^2 x^2
\over (1+s^2 x^2)^{1/2} }
\eqno (n5')
$$
where $\pi(x) = x^2/(1+e^{-x})$. For volatiles we have an analogous
expression:
$$
\Delta(2T_{00} - T) = {1 \over 2} \rho_{cr} \Omega_h z_{eq}^3 y^3
e_{3,v} (yd_\nu)
\eqno (5b)
$$
{\rm with}~~~
$$
e_{3,v} (s) = 2 \int_0^\infty dQ e^{-Q} 
{ 1/2 + s^2 Q \over (1+s^2 Q)^{1/2} }
\eqno (b5')
$$
In the gravitational field equations also the quantity $
e_4 = \int_0^\infty dx e^{-x} x \pi (x) \sigma_1 (x)$
is used, for massive  $\nu$'s ($\sigma_1$ is the first order
term in the spherical harmonic expansion of $\epsilon$). For volatiles
it ought to be replaced by
$$
e_{4,v} = \int_0^\infty dQ e^{-Q} \sqrt{Q} \sigma_1 (Q) ~.
\eqno (b6)
$$


Our code integrates the set of equations for the fluctuation evolution
in the synchronous gauge. We sped it up by truncating 
the hierarchical set of equations of free--streaming 
particles (photons after recombination, 
HDM and massless $\nu$'s), according to the
scheme suggested by Ma $\&$ Bertschinger (1996),
slightly modifying it because we 
don't use conformal time.
Results with truncation were compared with results
with a free number of harmonics and
we found that a truncation at the 8th harmonics is mostly sufficient.
Results presented here are however obtained with truncation
at the 24th harmonics. Models were evolved down to $z=0$.
We evaluated the transfer function for 24 scales 
ranging from $L=20000\, $Mpc to $L=0.53054\, $Mpc. Results from the public code
CMBFAST, yielding the transfer function for mixed models
(only with thermal hot component), were compared with the results of our 
algorithm, for a number of models and for scales fixed by CMBFAST.

Relative discrepancies between code outputs vary from less than $10^{-4}$,
on scales above $\sim 100\, $Mpc, to slightly 
above 3$\, \%$ in the worst cases, which occur for the hot component,
when $\Omega_h > 0.3$ and for scales below 2--3$\, $Mpc.
In such cases CMBFAST tends to give a transfer function slightly
smaller than our code. The typical discrepancy, however, keeps safely 
below $0.5\, \%$ and, therefore, we fully confirm the validity of CMBFAST.

Altogether the main source of error comes from
the analytic fit to the transfer functions, although the top discrepancy
at a single point keeps $< 1\, \%$ in the worst cases and its typical
value is $< 0.1\, \%$.

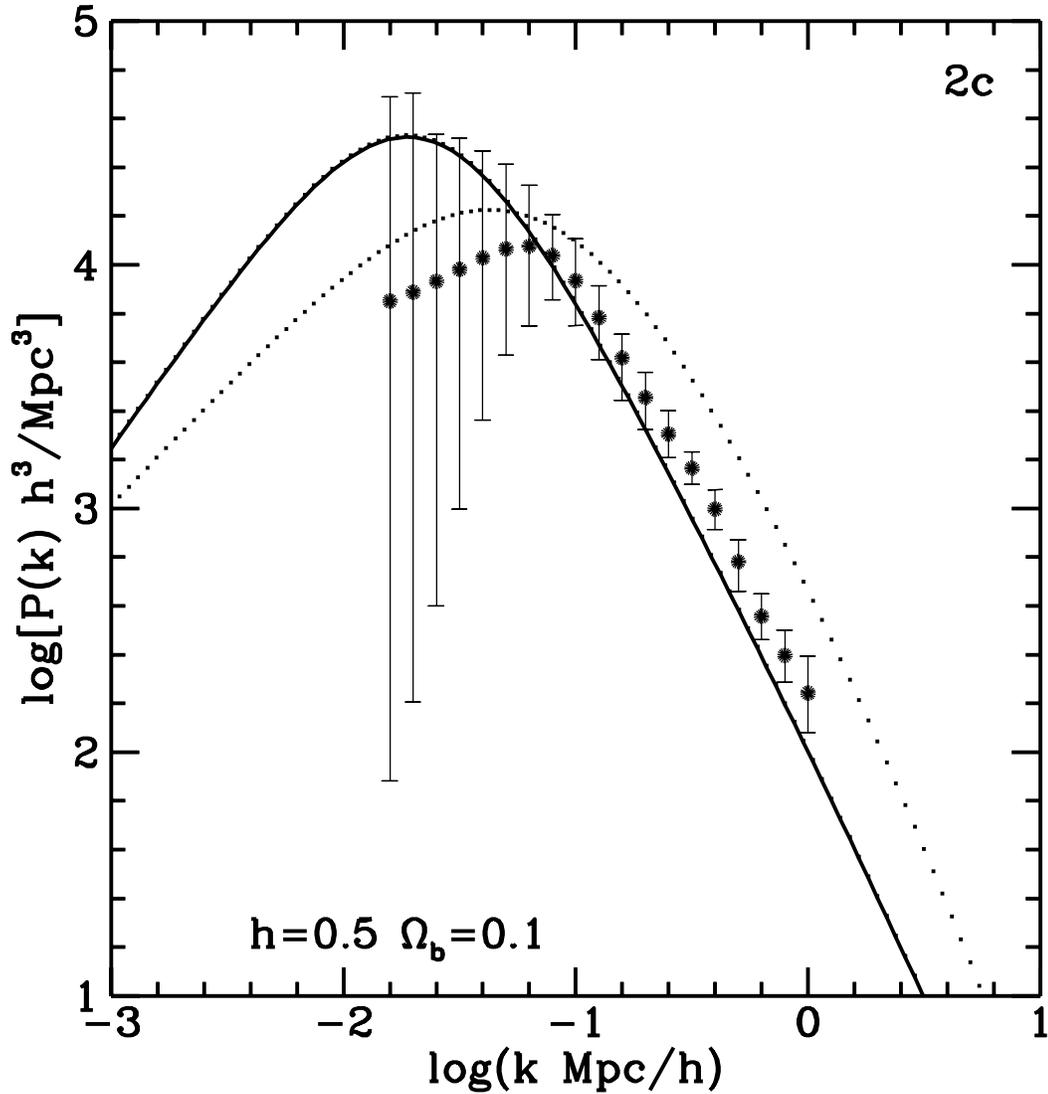
\end{document}